\documentclass{Interspeech2024}

\usepackage{pifont}
\usepackage{multirow}
\usepackage{float}
\usepackage{cleveref}
\usepackage{subcaption}

\newcommand{\method}{\textsc{NAST}\xspace}
\newcommand{\tsc}{\textsc{tSC}\xspace}
\newcommand{\newpara}[1]{\noindent {\bf #1}}
\interspeechcameraready
\usepackage{bm}
\usepackage{amssymb,amsmath,graphicx,bbm,color,booktabs}
\usepackage{amsopn}

\newcommand{\vp}{\bm{p}}

\newcommand{\vu}{\bm{u}}               
\newcommand{\vv}{\bm{v}}               
               
\newcommand{\vx}{\bm{x}}               
               
\newcommand{\vz}{\bm{z}}               

\renewcommand{\eqref}[1]{Eq.~(\ref{#1})}

\title{\method: Noise Aware Speech Tokenization for Speech Language Models}

\name{Shoval Messica and Yossi Adi}

\address{School of Computer Science and Engineering\\
The Hebrew University of Jerusalem, Israel}
\email{shoval.messica@mail.huji.ac.il}

\keywords{speech tokenization, speech language modeling}

\begin{document}

\maketitle

\begin{abstract} 
Speech tokenization is the task of representing speech signals as a sequence of discrete units. Such representations can be later used for various downstream tasks including automatic speech recognition, text-to-speech, etc. More relevant to this study, such representation serves as the basis of \emph{Speech Language Models}. In this work, we tackle the task of speech tokenization under the noisy setup and present \textbf{\method}: \textbf{N}oise \textbf{A}ware \textbf{S}peech \textbf{T}okenization for \emph{Speech Language Models}. \method is composed of three main components: (i) a predictor; (ii) a residual encoder; and (iii) a decoder. We evaluate the efficiency of \method considering several spoken language modeling tasks and show that \method is superior to the evaluated baselines across all setups. Lastly, we analyze \method and show its disentanglement properties and robustness to signal variations in the form of noise, reverberation, pitch-shift, and time-stretch. Code and pre-trained models are available at \url{https://github.com/ShovalMessica/NAST}.
\end{abstract}
\section{Introduction}
\label{sec:intro}

Self-supervised models have shown to be highly effective in extracting meaningful representations from raw speech signals~\cite{hubert, chen2022wavlm, baevski2020wav2vec, mohamed2022self}. Recently, the authors in~\cite{on_generative} demonstrated that such self-supervised representations can be used under the \emph{Generative Spoken Language Modeling} (GSLM) framework.

The GSLM pipeline typically starts with a self-supervised learning model that extracts continuous speech embeddings. These embeddings are then quantized into a discrete form, often using the k-means algorithm ~\cite{on_generative, kharitonov2021text, borsos2022audiolm}. A speech-language model is subsequently trained on these quantized units, which are finally converted back into raw audio through a unit-based neural vocoder. This framework was shown to be effective in modeling multiple levels of the speech utterance: prosody, content~\cite{on_generative, kharitonov2021text, borsos2022audiolm, kharitonov2023speak}, speech compression and enhancement~\cite{polyak2021speech, wang2023selm, erdogan2023tokensplit}, voice and emotion conversion~\cite{kreuk2021textless, maimon2023speaking}, spoken dialogue~\cite{nguyen2022generative}, and speech-to-speech translation~\cite{lee2021direct, popuri2022enhanced, lee2022textless, wang2023speech}. 

\begin{figure}[t!]
    \centering
    \includegraphics[width=0.4\textwidth]{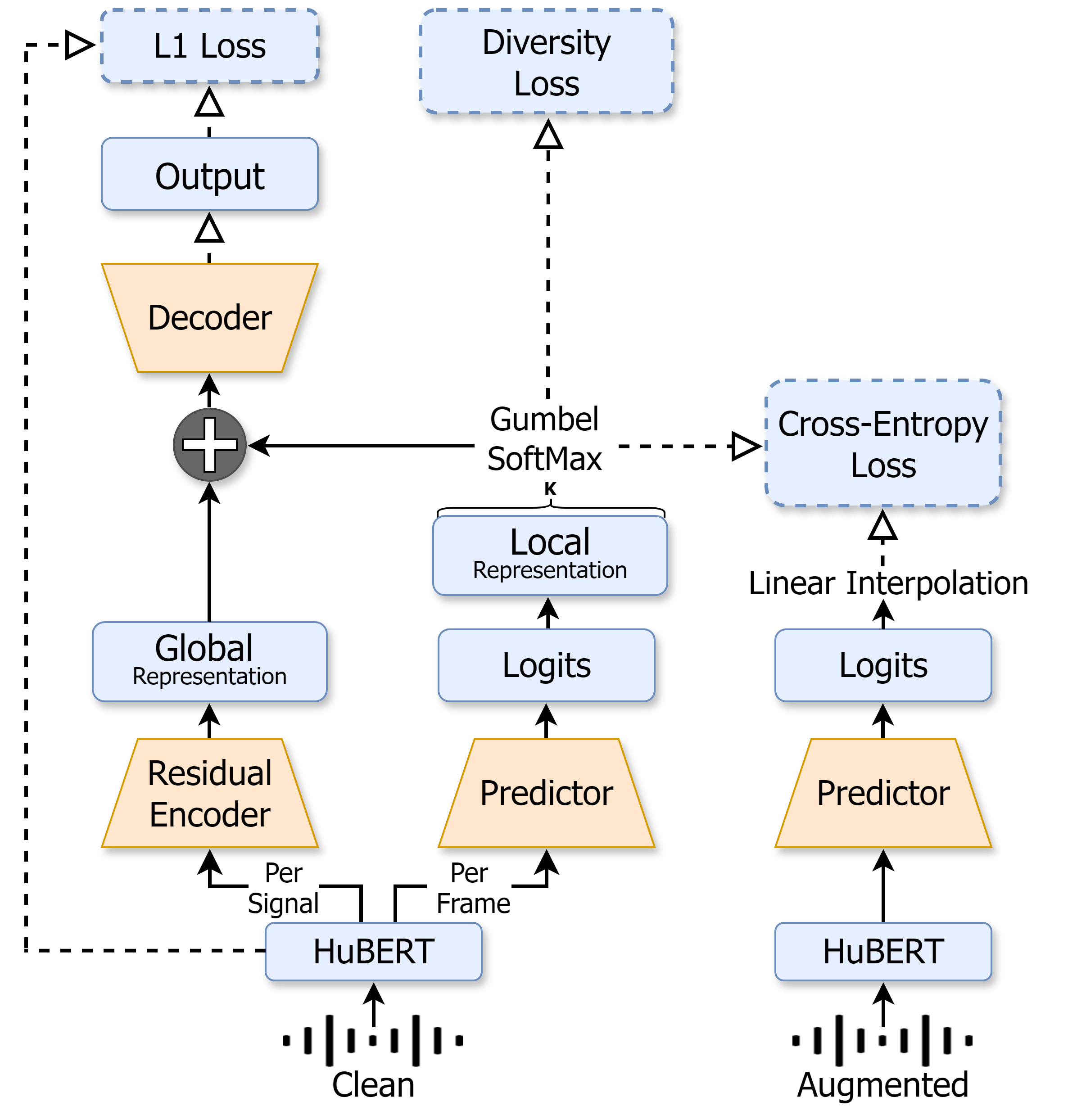}
    \caption{A high-level overview of \method. Clean and augmented signals are fed into the predictor to produce frame-wise logits. The clean logits undergo Gumbel sampling to become one-hot vectors for local representation. The residual encoder extracts a global representation from the clean signal, merged with local ones for decoder input to reconstruct the original signal embeddings. Augmented signal logits are aligned via linear interpolation for robustness enhancement, and diversity loss is applied over the one-hot vectors to ensure full unit usage.\label{fig:nast}
    \vspace{-0.5cm}}
\end{figure}

Despite their effectiveness, recent findings highlight a susceptibility of such techniques to acoustic variations that do not affect the linguistic content but greatly modify the output representation~\cite{gat-etal-2023-augmentation}, hence questioning their robustness and generalization. For instance, performing a time-stretch of less than 10\% of the speech utterance yields an edit distance of more than 40\%. The authors in~\cite{gat-etal-2023-augmentation} proposed an augmentation invariant speech tokenizer together with an objective evaluation metric to track progress in the field. Although providing impressive results, the method proposed by~\cite{gat-etal-2023-augmentation} is based on a teacher-student paradigm with k-means being the teacher. Hence, inherits k-means properties and bias.

To confront the above-mentioned issue, in this work, we propose a novel speech tokenization method named \textbf{\method}, which stands for \textbf{N}oise \textbf{A}ware \textbf{S}peech \textbf{T}okenization for Speech Language Models. \method is composed of three main components: (i) a predictor that maps the speech signal into local discrete representations. Such representation mainly captures local information in the form of phonemes or sub-phonemes; (ii) a residual encoder which predicts a single global representation for the whole sequence. This representation mainly captures global information such as speaker identification; and (iii) a decoder which outputs the original signal representation given both local and global representations. To improve invariance to signal variations we match the representations obtained from the predictor module of both clean and augmented speech signals. All modules are jointly optimized using several loss functions. Our hypothesis is that a representation that truly embodies the phonemic structure of speech will exhibit greater resilience, maintaining the spoken content even when faced with augmentations. Figure~\ref{fig:nast} visually depicts \method.

We evaluate \method's invariance to signal variations (i.e., time-stretch, pitch-shift, additive-noise, and reverberation), encoding capabilities (ABX), together with zero-shot sequence modeling evaluations, i.e., sWUGGY, sBLIMP~\cite{nguyen2020zero}, and Spoken StoryCloze~\cite{hassid2023textually}. Results suggest \method is comparable or superior to the evaluated baselines across all evaluation methods, considering various cluster numbers. We additionally analyze the learned representation considering speaker information and invariance to noise. These results shed light on the properties captured by the proposed speech tokenizer. 

\vspace{-0.1cm}
\section{Method}
\vspace{-0.1cm}
\label{sec:method}
The proposed method is comprised of three synergistic components: a frame-level Predictor $E_l$ outputting logits $\vp_{E_l}$ across discrete units, a Residual Encoder $E_g$ generating a global representation of the speech utterance, and a Decoder $D$ reconstructing the original representation given the concatenated representations. A visual description of \method can be seen in Figure~\ref{fig:nast}.

The input is an internal representation obtained from the 9th layer of a pre-trained HuBERT model~\cite{hubert}. Formally, we define the representation of the speech utterance as $\vx=\left(x_1, \dots, x_T\right)$ of $T$ frames. Our goal is to establish a mapping $E_l\left(\vx\right)=\vz=\left[z_1, \dots, z_T\right]$, where each $z_i \in \left[1 \dots k\right]$ is a categorical variable representing one of $k$ discrete units. Here, $E_l$ denotes the quantizer model we seek to learn. In parallel, we aim to learn a global representation $E_g\left(\vx\right)=\vu$, capturing global information about the input sequence.

\vspace{-0.1cm}
\subsection{Network architecture}
\vspace{-0.1cm}
\newpara{Predictor network.} The predictor network, $E_l$, takes $\vx$ as input and predicts a discrete distribution over the learned units at each time step, denoted as $\vp_{E_l}\left(\cdot\,|\,\vx,t\right)$.
That is, each frame $t\in T$ is associated with a $k$-dimensional logits vector in \( \mathbb{R}^{k}\). By employing a Gumbel-SoftMax operation on the logits, we derive a one-hot encoded vector $\textbf{1}_t$ in a differential manner, which we consider as a \emph{local representation}. In essence, the one-hot vector $\textbf{1}_t$ not only assigns the $t$-th frame to a specific unit but also contributes to the gradual shaping of each unit's meaning. Through exposure to diverse speech patterns and contexts, the predictor learns to refine the mapping such that each unit increasingly reflects consistent aspects of spoken content.

In order to sample from the predicted logits, we leverage the Gumbel-SoftMax reparametrization trick \cite{jang2016categorical}, enabling the model to simulate discrete choice within a differentiable framework.
Formally, the Gumbel-SoftMax distribution offers a continuous proxy for the categorical distribution. This continuity is achieved by introducing Gumbel noise to the logits associated with each category, followed by applying the SoftMax function:\begin{equation}
\begin{aligned}
\Vec{1}_t\left(i\right) = \frac{\exp\left(\frac{\log(\pi_i) + g_i}{\tau}\right)}{\sum_{j=1}^{k} \exp\left(\frac{\log(\pi_j) + g_j}{\tau}\right)}, \,\,\, \pi_{i}\triangleq\vp_{E_l}\left(i\,|\,\vx,t\right),
\end{aligned}
\end{equation}
where $\pi_i$ denotes the logit of the $i$-th unit, $g_i$ is a Gumbel noise sample, $k$ is the number of learned units, and $\tau$ is the temperature parameter that modulates the distribution's discreteness. A low $\tau$ yields a more discrete-like distribution, mimicking a one-hot encoded vector, thus facilitating the transition from continuous logits to a quasi-one-hot representation.

\newpara{Residual encoder.} In opposite to the predictor module, which operates at the frame level, the residual encoder $E_g$ aims to capture the global information within the signal. For that, we average the representations obtained by $E_g$ over time. Formally, for a speech utterance $\vx$, it's \emph{global representation} is given by: $\vu \triangleq E_g\left(\vx\right) = \frac{1}{T} \sum_{t=1}^{T} E_g\left(x_t\right)$.

This disentanglement facilitates the isolation of global attributes, thereby enabling the predictor to focus on modeling the local, content-driven nuances of speech.

\vspace{-0.1cm}
\subsection{Objective functions}
\vspace{-0.1cm}
The total objective function is a composite loss consisting of three key components: (i) Reconstruction loss, which aims to reconstruct the original HuBERT representation; (ii) Robustness loss, which ensures consistency of the model's tokenization in the presence of augmentations that do not alter the core spoken content; and (iii) Diversity loss, designed to promote the use of a wide range of units within the model.

\newpara{Reconstruction Loss.}
For each frame $t \in T$, the decoder $D$ reconstructs the original HuBERT embedding, given as input the concatenated local and global representations, denoted as $\textbf{1}_{t}$ and $\vu$, respectively. $\mathcal{L}_{\text{recon}}$ is defined as:
\begin{equation}
\begin{aligned}
\mathcal{L}_{\text{recon}}\left(E_l, E_g, D\,;\, \vx\right) = \frac{1}{T} \sum_{t \in T} L_1\Big( D\big(\textbf{1}_{t} \oplus \vu\big), x_t \Big)
\end{aligned}
\end{equation}

\newpara{Robustness Loss.} We wish to maintain consistent tokenization of a clean speech utterance $\vx$ across various augmentations that preserve the spoken content. In line with~\cite{gat-etal-2023-augmentation}, we adopt an alignment-based strategy to enhance the invariance of \method, utilizing a wide range of augmentations.

Formally, consider a clean speech utterance $\vx$ of $T$ frames and its augmented version $\Vec{\tilde{x}}$ of $T'$ frames. Initially, both are fed into the predictor $E_l$, which outputs $T$ and $T'$ logit vectors for each, respectively. From the logit vectors of the clean signal, we derive $T$ one-hot vectors, each representing the model's chosen unit for the corresponding frame, forming the ``ground truth'' in this context. To address the temporal discrepancies induced by augmentations, we use linear interpolation to adjust the number of logit vectors from the augmented signal $\Vec{\tilde{x}}$ to align with the number of one-hot vectors from the clean signal $\vx$:
\begin{equation}
\vp'_{E_l}(\cdot|\Vec{\tilde{x}}) = \text{Interpolate}\left(\vp_{E_l}(\cdot|\Vec{\tilde{x}}), T\right)
\end{equation}
Next, we make a frame-wise comparison between the aligned augmented logits, $\Vec{p}'_{E_l}\big(\cdot|\Vec{\tilde{x}}, 1\big) \cdots \vp'_{E_l}\big(\cdot|\Vec{\tilde{x}},T\big)$, and the clean signal's one-hot vectors, $\Vec{1}_1, \ldots, \Vec{1}_T$, as follows:
\begin{equation}
\begin{aligned}
\Vec{v}_t\left(i\right) \triangleq \frac{\exp\left(\vp'_{E_l}\left(i\,|\,\Vec{\tilde{x}}, t\right)\right)}{\sum_{j=1}^{k} \exp\left(\vp'_{E_l}\left(j\,|\, \Vec{\tilde{x}}, t\right)\right)} 
\end{aligned}
\end{equation}
\begin{equation}
\begin{aligned}
\mathcal{L}_{robust} = \sum_{t=1}^{T} \overbrace{-\sum_{i=1}^{k}
\Vec{1}_{t}\left(i\right)\cdot \log\left( \Vec{v}_t\left(i\right) \right)}^{\text{CrossEntropy}(\vv_t,\Vec{1}_{t})}
\end{aligned}
\end{equation}
This comparison iteratively refines the unit selection, encouraging the model to yield similar distributions for both augmented and original signals, thus enhancing augmentation invariance.

\newpara{Diversity Loss.}
To address a significant training challenge, where the model tends to ``saturate'' by favoring a limited set of units, we incorporate a Diversity Loss, drawing inspiration from the work of~\cite{baevski2020wav2vec}. Applied to the clean speech utterance $\vx$, we define the diversity loss as follows:
\begin{equation}
    \mathcal{L}_{diversity} = \frac{1}{k} \sum_{i=1}^{k} \bar{p}_i \log(\bar{p}_i),\,\, \bar{p}_i = \frac{1}{T} \sum_{t=1}^{T} \textbf{1}_{t}(i)
\end{equation}
Here, \(\bar{p}_i\) quantifies the average selection rate of the $i$-th unit by the predictor over the $T$ frames within the clean utterance $\vx$. 

Overall, the final objective function is computed as a weighted sum of the three terms:
\begin{equation}
\mathcal{L}_{total} = \mathcal{L}_{recon} + \lambda_{1} \cdot \mathcal{L}_{diversity} + \lambda_{2} \cdot \mathcal{L}_{robust}
\end{equation}
where $\lambda_{1}$ and $\lambda_{2}$ are tunable hyperparameters.
\vspace{-0.1cm}
\section{Experimental Setup}
\vspace{-0.1cm}
\label{sec:setup}

In all setups we consider the base version of HuBERT~\cite{hubert}, using the $9$th' layer, operating at $50$Hz, using $50$, $100$, and $200$ clusters.
We optimize, compare, and analyze the performance of various SpeechLMs (ULMs) learned over the discrete units, considering different setups. The training of both the ULMs and the quantizers is efficiently executed on a single NVIDIA A$6000$ GPU.
We followed a similar setup as in \cite{gat-etal-2023-augmentation}, where we subject the speech signals to a suite of augmentations each chosen for its potential to simulate real-world signal variability and thereby refine the learned units' robustness. Specifically, we consider: (i) \textbf{time-stretch} using a Phase Vocoder method \cite{karrer2006phavorit} to stretch or shrink the time domain signal in the range $[0.8, 1.2]$ without changing the pitch; (ii) \textbf{pitch-shifting} the speech signal by four semitones using the resampling method over the time-stretched signal~\cite{karrer2006phavorit}; (iii) \textbf{reverberation} following similar setting as~\cite{chazan2021single} simulated via the pyroomacoustics~\cite{scheibler2018pyroomacoustics} audio room simulations package; and (iv) \textbf{noise injection} using a randomly sampled Signal-to-Noise Ratio (SNR) in the range of $[5, 15]$. Background noises are sampled from the Deep Noise Suppression (DNS) challenge~\cite{reddy2020interspeech} which includes a diverse set of noise types from AudioSet~\cite{gemmeke2017audio}, Freesound,~\cite{font2013freesound}, and Demand~\cite{thiemann_joachim_2013_1227121}.

\vspace{-0.1cm}
\subsection{Model \& Hyperparameters}
\vspace{-0.1cm}
\newpara{Speech encoding.} We use a learning rate of $1e-4$ and a batch size of $16$, employing the Adam optimizer~\cite{adam}. The architecture of our network features Conformer blocks \cite{gulati2020conformer}, each followed by a projection layer. The projection layer sizes for the decoder $D$ and the residual encoder $E_g$ are set to $768$ and $256$, respectively. The size of the predictor $E_l$ projection depends on the number of units, $k$. Empirically determining the optimal configuration for each unit count involves assessing the number of layers, attention heads, kernel size, and feed-forward network (FFN) dimension. 
All quantizers are trained using the LS 960h dataset. We employed a weighted loss, where $\lambda_1$ and $\lambda_2$ denote the weights assigned to the diversity and robustness losses, respectively. We initiated the hyper-parameters to $\lambda_1 = 1$ and $\lambda_2 = 0.005$. We observed a sensitive interaction between the losses, where an increase in $\lambda_2$ tended to reduce the unit diversity. To address this, we carefully tuned both $\lambda_1$ and $\lambda_2$ upwards, ensuring that any adjustments were made in a controlled manner to prevent one loss from overshadowing the other.

\newpara{unit-LM:} For each number of clusters $k$, we train two types of uLM: one operating over our units and the other over units obtained via k-means quantization. The k-means quantizers are derived from the textless-lib \cite{kharitonov2022textless}. These uLMs are based on the \texttt{transformer\_lm\_big} architecture implemented in \texttt{fairseq}~ \cite{ott2019fairseq}, where each sample within the batch contains up to $4,096$ units. During training, these models operate as causal language models on sequences of deduped units as in~\cite{on_generative}. All language models are trained on a ``clean'' $6$k-hour sub-sample of LibriLight~\cite{Kahn_2020}.
\begin{table}[t!]
\caption{Augmentation invariance results. UED is reported for \method, k-means, and Gat et al., considering additive noise, time-stretch, reverberation, and pitch-shift. Results are reported for $50$, $100$, and $200$ units. \label{tab:ued_scores}}
\centering
\resizebox{.47\textwidth}{!}{
\begin{tabular}{@{}llcccc@{}}
\toprule
\multirow{2}{*}{\textbf{Units}} & \multirow{2}{*}{\textbf{Method}} & \multicolumn{4}{c}{\textbf{Augmentation UED }$\downarrow$} \\
\cmidrule(lr){3-6}
 & & \textbf{Noise} & \textbf{Time-Stretch} & \textbf{Reverb} & \textbf{Pitch Shift} \\
\midrule
\multirow{3}{*}{50} & K-Means & 29.74 & 39.61 & 28.25 & 44.33 \\
 & Gat, et al~\cite{gat-etal-2023-augmentation} & 24.67 & 26.89 & 19.89 & 30.22 \\
  & \textbf{NAST} & \textbf{9.51} & \textbf{17.26} & \textbf{9.82} & \textbf{16.47} \\
\midrule
\multirow{3}{*}{100} & K-Means & 31.38 & 41.97 & 30.42 & 48.68 \\
 & Gat, et al~\cite{gat-etal-2023-augmentation} & 25.06 & 29.72 & 21.31 & 32.84 \\
 & \textbf{NAST} & \textbf{10.82} & \textbf{17.45} & \textbf{10.35} & \textbf{18.74} \\
\midrule
\multirow{3}{*}{200} & K-Means & 33.34 & 45.59 & 32.89 & 53.14 \\
 & Gat, et al~\cite{gat-etal-2023-augmentation} & 26.76 & 32.99 & 22.94 & 36.45 \\
 & \textbf{NAST} & \textbf{11.88} & \textbf{21.36} & \textbf{13.86} & \textbf{22.97} \\
\bottomrule
\end{tabular}}
\end{table}

\begin{table*}[t!]
\caption{Modeling and encoding evaluation. We report results for \method, k-means, and Gat et al.~\cite{gat-etal-2023-augmentation}. Results of ~\cite{gat-etal-2023-augmentation} were taken from the paper as no code / pre-trained models were publicly available.}
\centering
\captionsetup{justification=centering} 
\begin{tabular}{@{}llccccccccccc@{}} 
\toprule
\multirow{2}{*}{\textbf{Units}} & \multirow{2}{*}{\textbf{Method}} & 
\multicolumn{2}{c}{\textbf{sWUGGY} $\uparrow$} & \multicolumn{2}{c}{\textbf{sBLIMP} $\uparrow$} & 
\multicolumn{2}{c}{\textbf{\tsc} $\uparrow$} & \multicolumn{2}{c}{\textbf{ABX (clean)} $\downarrow$} & \multicolumn{2}{c}{\textbf{ABX (other)} $\downarrow$} \\ 
& & clean & aug & clean & aug & clean & aug & within & across & within & across \\ 
\midrule
\multirow{3}{*}{50} & K-Means & 67.48 & 62.61 & 52.42 & 51.81 & 66.27 & 59.95 & 7.52 & 8.90 & 9.84 & 13.50 \\
 & Gat, et al~\cite{gat-etal-2023-augmentation} & 67.59 & - & 53.68 & - & - & - & 6.63 & 7.55 & 9.53 & 12.14 \\
  & \textbf{NAST} & 67.14 & 63.69 & 54.34 & 52.78 & 64.51 & 59.32 & 5.85 & 6.74 & 7.77 & 10.25 \\
\midrule
\multirow{3}{*}{100} & K-Means & 67.75 & 64.46 & 51.96 & 51.82 & 67.18 & 60.55 & 6.37 & 7.72 & 8.4 & 12.29 \\
 & Gat, et al~\cite{gat-etal-2023-augmentation} & 68.20 & - & 53.12 & - & - & - & 5.39 & 6.22 & 7.46 & 10.20 \\
 & \textbf{NAST} & 73.35 & 70.01 & \textbf{55.86} & 54.72 & 64.13 & 61.78 & \textbf{5.20} & \textbf{5.90} & \textbf{6.92} & \textbf{8.73} \\
\midrule
\multirow{3}{*}{200} & K-Means & 71.88 & 66.56 & 52.43 & 51.15 & \textbf{67.55} & 58.04 & 5.99 & 7.14 & 8.23 & 11.51 \\
 & Gat, et al~\cite{gat-etal-2023-augmentation} & 70.68 & - & 54.91 & - & - & - & 5.19 & 6.00 & 7.18 & 9.70 \\
 & \textbf{NAST} & \textbf{76.42} & \textbf{71.79} & 55.62 & \textbf{55.07} & 66.70 & \textbf{64.45} & 5.47 & 6.22 & 7.18 & 9.43 \\
\bottomrule
\end{tabular}
\label{tab:metrics}
\end{table*}

\vspace{-0.1cm}
\section{Results}
\vspace{-0.1cm}
We evaluate \method across two axes: (i) speech encoding evaluation; and (ii) sequence modeling evaluation. We additionally provide an analysis to better highlight the properties of the proposed quantizer. We compare \method to the commonly used k-means method~\cite{on_generative, borsos2022audiolm} and the method proposed by ~\cite{gat-etal-2023-augmentation}, considering $50$, $100$, and $200$ clusters.

\vspace{-0.1cm}
\subsection{Speech encoding evaluation}
\vspace{-0.1cm}
We evaluate speech encoding considering both its phonetic discriminative capabilities using the ABX metric together with invariance to signal variations. 

\newpara{Augmentation invariance:} We utilized the Unit Edit Distance UED metric proposed by \cite{gat-etal-2023-augmentation} to evaluate the robustness of \method to signal variations. This metric, based on the Levinstein Distance \cite{yujian2007normalized}, measures the dissimilarity between the clean and augmented signals in terms of de-duplicated units. Ideally, a perfect spoken language quantizer would obtain a zero distance after deduplication. Table~\ref{tab:ued_scores} presents the UED scores across various augmentations and $k$ values. Notably, our method consistently outperforms previous works by a significant margin, indicating its superiority in handling these challenges.

\newpara{ABX:} The ABX task examines the discriminative phonetic abilities of the representation. It involves a pair of words differing by a single phoneme and a reference test word sharing a phoneme with one of the pair. It assesses whether the test phoneme is closer in representation to the correct or incorrect phoneme, expecting a shorter distance to the correct one.
The ABX task is conducted in two setups: 'within' and 'across'. 'Within' is evaluated on input data from the same speaker, while 'across' is evaluated on input data from different speakers. Table~\ref{tab:metrics} shows the ABX results for both. We observed that the ABX-clean score was consistently on par with or exceeded the benchmarks established by other methods. Importantly, our approach demonstrated significant and steady enhancements in the more challenging 'other' split, which is distinguished by recordings that feature background noise and a variety of accents. 

\vspace{-0.1cm}
\subsection{Sequence modeling evaluation} 
\vspace{-0.1cm}
Under the modeling evaluation, we consider both the zero-resource speech metrics together with \emph{Spoken Story Cloze}. To highlight the invariance of \method to signal variations, we report results for two data versions: clean signals and noisy signals. The clean version is the original benchmark, while the noisy is an augmented version of the speech utterance.

\newpara{Zero resource speech:} We start by evaluating \method using the zero resource speech metrics~\cite{nguyen2020zero}, i.e., sWUGGY, and sBLIMP. The sWUGGY metric requires detecting the real word from a pair of short utterances such as 'brick' vs. 'blick.' Similarly, sBLIMP requires detecting the syntactically correct sentence from a pair of sentences. In both metrics, detection is done by comparing the probabilities of both sequences. As summarized in Table~\ref{tab:metrics}, \method exhibited significant sWUGGY and sBLIMP improvements across most setups. As expected, these improvements are kept both under the clean and augmented versions. 

\newpara{Spoken StoryCloze:} Next, we adapted the topic version of the Spoken StoryCloze Benchmark (\tsc)~\cite{hassid2023textually} to assess our Unit Language Models' ability to understand narrative continuity and capture intricate nuances. This benchmark involves distinguishing the correct ending from an adversarial one in a set of $4,000$ five-sentence commonsense stories. Results are summarized in Table~\ref{tab:metrics}. When considering the clean version, \method provides inferior performance to the k-means alternative, however, when considering the noisy version \method consistently outperforms the k-means method.

\begin{figure}[t!]
    \centering
    \begin{subfigure}{0.4\textwidth}
        \centering
        \includegraphics[width=\textwidth, trim={0 40 0 30}, clip]{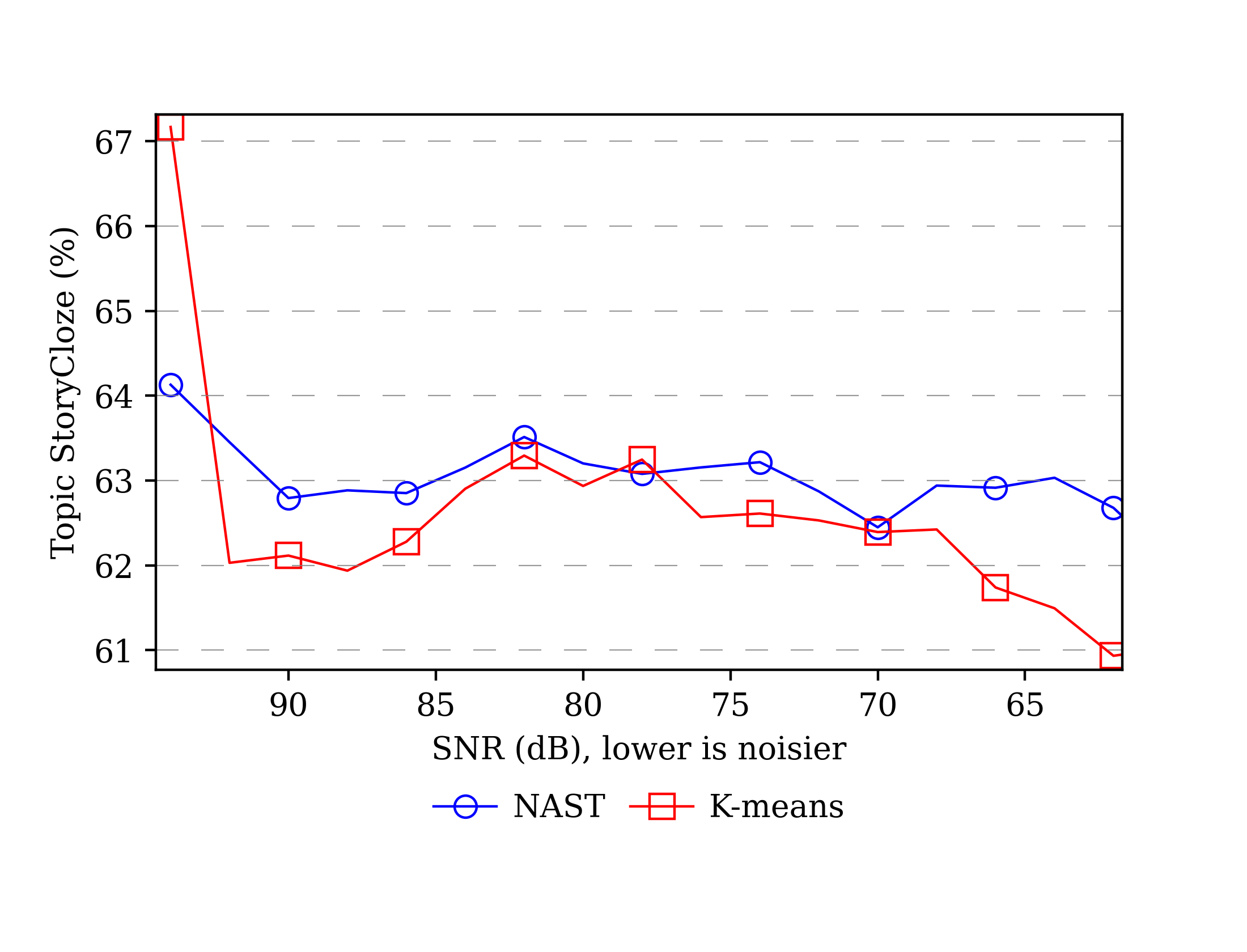}
        \caption{\label{fig:tsc_noise}}
    \end{subfigure}    
    \begin{subfigure}{0.4\textwidth}
        \centering
        \includegraphics[width=\textwidth, trim={0 5 0 15}, clip]{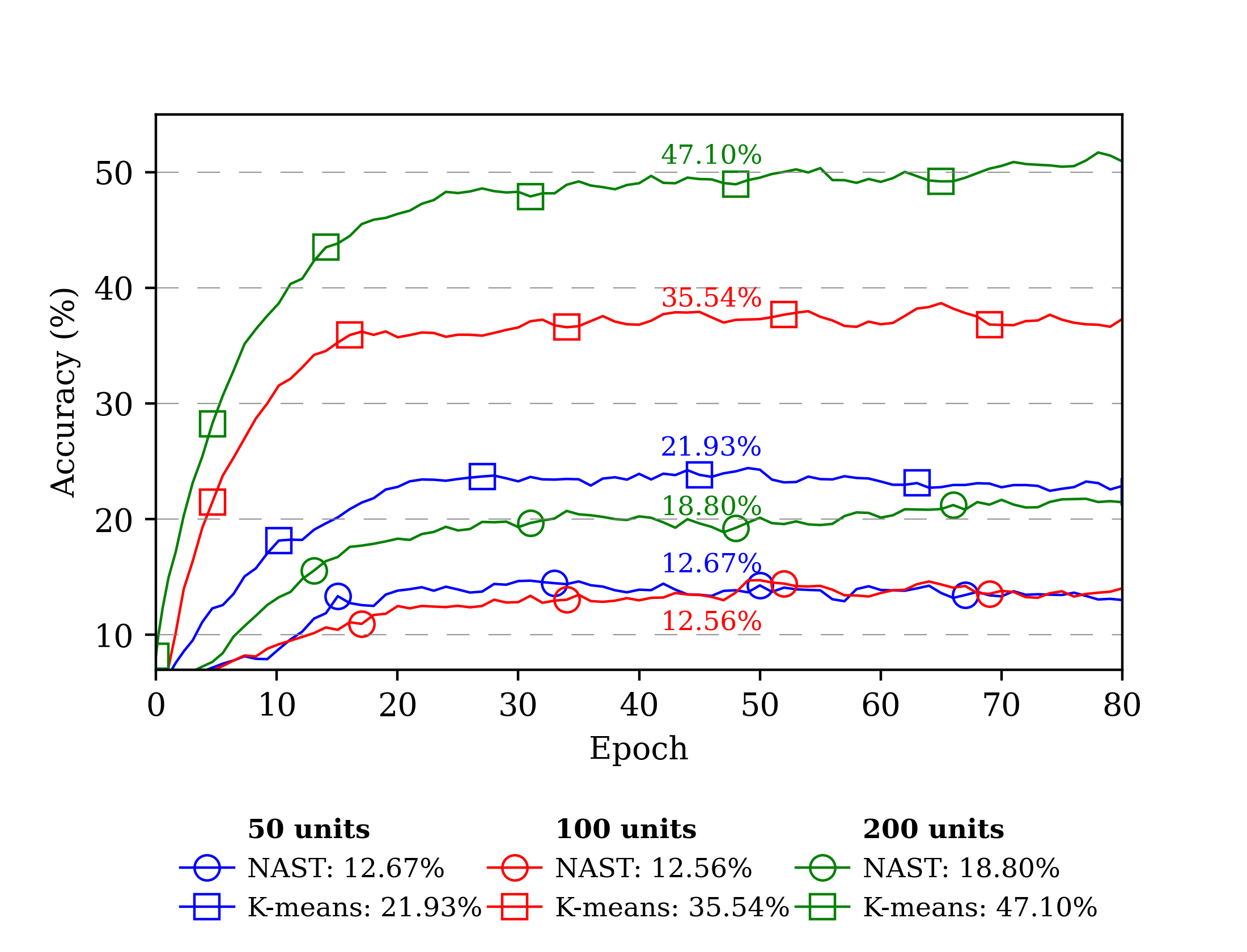}
        \caption{\label{fig:speaker_local}}
    \end{subfigure}
    \caption{(a) \tsc performance as a function of noise levels. Results are reported for both \method and k-means using $100$ clusters. (b) Speaker Probing for Local Representation: Classifiers trained for 100-epoch on LibriSpeech 'dev-clean'. 
    % Mean accuracy across epochs marked on the curves.
    \label{fig:analysis}\vspace{-0.6cm}}
\end{figure}

\vspace{-0.1cm}
\subsection{Analysis}
\vspace{-0.1cm}
\newpara{Noise invariance.} To analyze the models' sensitivity to augmentations, we progressively increase the level of augmentations and measure the impact on model performance. We measure the \tsc performance as a function of noise levels, measured in SNR. As shown in Figure \ref{fig:tsc_noise}, although the k-means alternative performs better under clean samples, as we increase the noise levels its performance rapidly decreases, this is in contrast to \method which provides much more stable results.

\newpara{Speaker probing.} Lastly, we analyze the differences between global and local representations in capturing global information, specifically, we consider speaker information. For that, we perform speaker probing~\cite{kharitonov2022textless, adi2019reverse}. This task assesses the model's ability to identify an anonymized speaker ID based on their utterances, thus reflecting the extent to which speaker-specific information is retained in the representations. We compared \method's local representations against k-means clustering, in Figure~\ref{fig:speaker_local}. Results suggest the local representation obtained by \method contains less speaker information compared to the k-means one across all setups, whereas as we increase the number of clusters the gap increases. In contrast, when investigating the \textbf{global representations}, \method achieves high accuracies, with \textbf{92.73\%}, \textbf{98.24\%}, and \textbf{97.67\%} for \textbf{50}, \textbf{100}, and \textbf{200} units, respectively. These results emphasize the residual encoder's efficacy in differentiating between local and global information.
\section{Conclusion}
\label{sec:con}

This study introduces a novel speech tokenization method that enhances the robustness and performance of GSLM models. By adopting a dynamic strategy to map and refine speech units, our method departs from the traditional k-means clustering, introducing an innovative way to tokenize spoken language. Supported by comprehensive evaluations, our model demonstrates superior performance in handling diverse acoustic variations. Looking ahead, we intend to develop advanced and hierarchical tokenization techniques, enhancing our understanding and modeling of spoken language models.

\vspace{0.0cm}
\noindent \textbf{Acknowledgements} This research work was supported by ISF grant 2049/22.

\bibliographystyle{IEEEtran}
\bibliography{mybib}

\end{document}